\documentclass{article}

\usepackage{arxiv}

\usepackage[utf8]{inputenc} 
\usepackage[T1]{fontenc}    
\usepackage{hyperref}       
\usepackage{url}            
\usepackage{booktabs}       
\usepackage{amsfonts}       
\usepackage{nicefrac}       
\usepackage{microtype}      
\usepackage{lipsum}
\usepackage{graphicx}
\usepackage{indentfirst}
\title{Bistable Probabilities:
A Unified Framework for Studying Rationality and Irrationality in Classical and Quantum Games}

\author{
  Shahram Dehdashti\\
  School of Information Systems\\ Queensland University of Technology\\ Brisbane, Australia\\
  \texttt{shahram.dehdashti@qut.edu.au}
   \And
 Lauren Fell \\
School of Information Systems\\ Queensland University of Technology\\ Brisbane, Australia\\
  \texttt{13.fell@qut.edu.au} 
   \AND
    Abdul Karim Obeid \\
School of Information Systems\\ Queensland University of Technology\\ Brisbane, Australia\\
  \texttt{abdul.obeid@hdr.qut.edu.au} 
   \AND
    Catarina  Moreira \\
School of Information Systems\\ Queensland University of Technology\\ Brisbane, Australia\\
  \texttt{catarina.pintomoreira@qut.edu.au} 
   \AND
   Peter Bruza  \\
   School of Information Systems\\ Queensland University of Technology\\ Brisbane, Australia\\
  \texttt{p.bruza@qut.edu.au} }

\begin{document}






\maketitle
\begin{abstract}
This article presents a unified probabilistic framework that allows both rational and irrational decision making to be theoretically investigated and simulated in classical and quantum games.
Rational choice theory is a basic component of game-theoretic models, which assumes that a decision maker chooses the best action according to their preferences.
In this article, we define irrationality as a deviation from a rational choice. 
Bistable probabilities are proposed as a principled and straight forward means for modeling (ir)rational decision making in games. Bistable variants of classical and quantum  Prisoner’s Dilemma, Stag Hunt and Chicken are  analysed in order to assess the effect of (ir)rationality on agent utility and Nash equilibria.
It was found that up to three Nash equilibria exist for all three classical bistable games and maximal utility was attained when agents were rational.
Up to three Nash equilibria exist for all three quantum bistable games, however, utility was shown to increase according to higher levels of agent irrationality.
\end{abstract}


\section{Introduction}

Classical game theory is a mathematical framework that aims to understand the interactions of decision makers and their joint outcomes.
Decision makers include people, organizations, animals and intelligent machines.
Rational choice theory (RCT) is a basic component of game-theoretic models, which assumes that a decision maker chooses the best action according to their

\maketitle
\noindent
preferences.
Under certain assumptions, preferences are well behaved, and can therefore be represented as utility functions. Rationality then equates to maximizing the decision maker's utility, or finding the maximum value of the decision maker's utility function \cite{bicchieri:2004}.

Economists, sociologists, and other adherents of RCT typically view human behavior to be rational when it involves the maximization of the decision maker's expected utility.
In a somewhat weaker form, rationality is redefined in terms of choice consistency: The decision maker with a given set of preferences is rational to the extent that their choices are internally consistent with one another \cite{Hodgson:2012}. 

Compared to economic theory, the 
cognitive and psychological 
sciences show a much higher 
variability in their 
conceptualizations of rationality. 
Some scholars adhere to the 
postulates of RCT, though with some adaptations. These approaches 
include risk-weighted expected 
utility maximization, which 
captures the decision maker's 
attitudes toward risk \cite{buchak:2013},
the notion that rationality is local to a decision-making context \cite{vlaev:2018}, the rational analysis of 
cognition which assumes a function describing optimal behaviour  \cite{chater:oaksford:2000}, and models of the maximization of subjective expected utility under cognitive and task constraints \cite{howes:lewis:vera:2009}. 
All these approaches are somewhat aligned with RCT because they are based on  constrained maximization of a target function, which is determined by the decision maker's preferences.

Kahneman, Wakker, and Sarin \cite{kahneman1997back} step out of the bounds of RCT as they differentiate decision utility, stemming from RCT's assumptions of rationality and consistency of preferences, from experienced utility, which focuses on the predicted hedonic reward from an action. This reward can be immediate or remembered, and is not restricted to personal gain. In their review, Kahneman and Thaler \cite{kahneman2006anomalies} find multiple examples of biases in predicting future experience of reward, affecting the maximisation of utility.

In short, the interplay between rationality and decision making has been broadly debated in literature covering economics, cognitive  and social psychology.
Our intention is not to contribute to this debate.
Rather, our intention is to highlight that however rationality is defined, the decision maker, for whatever reason, may deviate from a rational choice. 
We term such a deviation ``irrationality". What then are the consequences for game theory?
This article aims to address  this question by introducing a deformed probability; a so-called ``bistable probability", as a formal means of systematically investigating deviations from a rational choice.

The bistable probability framework is used to theoretically analyse three well-studied games in the literature, i.e. Prisoner’s Dilemma (PD), Stag Hunt (SH) and Chicken Game (CG). By using these examples, we analyse how irrationality leads to additional Nash equilibria (NE) when compared to rational versions of the games. Using an example sourced from the literature, the effect of irrationality on the relationship between the benefit to cost ratio and cooperation is also investigated.

  Quantum game theory is a natural generalization of game theory \cite{eisert1999quantum,iqbal2002quantum,iqbal2016equivalence}. In quantum games, strategies are defined by  Hermitian operators that act on the initial states of players. The strategy space can be extended by considering the superposition of strategies and entanglement of players' choices. A player's rational decision in a quantum game corresponds to a sharp projection in Quantum Mechanics (QM). Positive operator value (POV) projections are un-sharp projections \cite{busch1986unsharp,spekkens2007evidence} that serve to extend a complex Hilbert space. We will show that bistable operators are equal to un-sharp projections, and can therefore be used to describe deviations from rationality.        Further, we demonstrate that POV projections from quantum mechanics create a suitable framework to model irrationality across classical and quantum games. 
Therefore, this article proposes POV projections as a general framework for both classical and quantum games involving both rational and irrational agents.
Through analysis of this framework, we show that agent utility decreases with the presence of irrationality in classical games but that the converse occurs in quantum irrational games.
Finally, we show that non-factorisable strategies are naturally obtained when we consider a bistable parameter in a simulated quantum game \cite{brunner2013connection,iqbal2016equivalence,alonso2019collective}. We indicate that this model can provide a framework by which cooperation in non-cooperative games can be modelled.\\
\indent The rest of the paper is organized as follows: In 
section \ref{sec2}, we  model irrationality using POV projections derived from QM. In section \ref{sec3}, a framework is suggested for irrational classical games; we study three different games: PD, SH and CG. In section \ref{sec4}, by using bistable Kraus projections, we study the three above mentioned games in a quantum setting. Section \ref{sec5} describes details of the mathematical framework. Finally, section \ref{sec6} is devoted to some conclusions and directions for future research. 
\section{Bistable probabilities}\label{sec2} 
The preceding section describes the manifold ways in which rationality is viewed. In one conceptualisation of rationality, rational agents might be expected to defect in one-shot Prisoners' Dilemma games, as this choice maximizes monetary gain. In another way, a rational agent may be expected to cooperate based on factors that include social \cite{capraro2014heuristics, moisan2018not}, moral \cite{dana2007exploiting}, personality \cite{balliet2009social}, or sociocultural  preferences\cite{rand2016social}, which may promote altruism over self-gain. The focus of the present paper is to model deviations from rationality, however it may be viewed. To do this, we consider an agent to be in two minds: one rational, one irrational. One can consider this in terms of having two sets of preferences inconsistent with one another. We map this intuition onto a Dual Process account of human decision making in order to clarify it.
 
Dual process theories are prominent and widespread within many fields of psychological science \cite{gladwin:figner:2015}. Whilst there are multiple variations of these theories, all have in common the thesis that two processes are employed in human decision-making: System 1 and System 2. System 1 produces an intuitive approach to decision making which is often characterized by heuristics and other mental shortcuts \cite{kahneman2011thinking}. System 1 is fast, requires few cognitive resources and little effort \cite{evans2003two}. System 2 involves controlled analytic thought, and is often considered to be logical, however requires conscious activation and is a significant drain on cognitive resources \cite{evans2003two}.

The preferences associated with each of these systems may be inconsistent with each other. For example, studies \cite{rand2012spontaneous, rand2014social, isler2018intuition, cone2014time} have found that, under certain conditions \cite{capraro2019dual}, decisions of cooperation tend to coincide with System 1, or intuitive, processing, while further deliberation may lead to decisions of defection. The Social Heuristics Hypothesis \cite{rand2014social} was proposed as a means to explain cooperation and predicts its prevalence when intuitive (System 1) thinking is employed. Whilst there is evidence to support the claim that intuition favours cooperation, this appears to be moderated by other  factors, such as gender differences \cite{rand2016social}. In addition to this, there are a range of methodological factors present in the study of dual processes that may also affect results (see review by \cite{capraro2019dual}). Nevertheless, it is clear that there are conflicting preferences across System 1 and System 2 which may place an agent in a game in two minds when making a choice. 

To position this in terms of rationality, consider a situation in which the preferences of an individual employing their deliberative, System 2 process promote the decision to defect. Their expected utility would be higher if they chose to defect than to cooperate. To defect in this specific instance, would therefore be the rational choice. However, this individual’s intuition (System 1) may promote a decision to cooperate. Employing System 1 may see the expected utility to be higher if the individual chose to cooperate rather than defect. In this case, cooperating could be considered to be the rational choice. When there is disagreement between System 1 and System 2, and when the preferences between the two systems are inconsistent, we can consider the potential for irrationality because the inconsistency may lead to a deviation from a rational choice. 

Although the bistable probability framework is agnostic to the order of System 1 and 2 processes, as well which system is deemed the rational one,
we can take the following example for ease of exposition. If System 2, and monetary gain, is seen to be the rational process and choice in a certain situation, we introduce irrationality as the deviation from that choice if System 1 preferences conflict with System 2. We introduce the bistable parameter $k$ to represent the level of disagreement between System 1 and System 2 in a similar way as in \cite{dehdashti2020irrationality}. It should be noted at this point that we can just as easily flip the explanation under the assumption that System 1 produces a choice that maximises the agent’s utility, and where System 2 is seen to deviate from this by a factor defined by $k$. In terms of the interaction between System 1 and System 2, there are various views about how this might occur, for example, in parallel or with System 1 acting as a default for which System 2 can intervene \cite{evans2007resolution}. The model proposed in this paper is agnostic towards the order or dynamics of System 1 and System 2 processes and describes only the deviation of one from the other. In this way, we present the bistable parameter as a means of systematically investigating irrationality in games.

\indent Inspired by \cite{costello2014surprisingly,costello2016people}, bistable probabilities are founded on an event space featuring two probabilities associated with a given event $E$: $p(E)$ and $p_k(E)$ where $p_{k}(E)=1-p-k+2kp$ and $p_{k}(\neg E)=p+k-2kp$. 
In terms of the previously mentioned ``two minds" metaphor, bistable probabilities can be interpreted as follows: The rational ``mind" (System 1 or System 2, depending on the prescribed theory of rationality) would make a decision $E$ with probability $p(E)$, where $p$ may be a classical or quantum probability depending on whether the game is classical or quantum. 
The preferences of the irrational ``mind" may deviate from the rational decision with probability $p_k(E)$.
This allows irrational decision making to be modelled as a deviation from the probability ascribed to a rational decision.
The deviation is modelled by the parameter $k$: 
When $k=1$, the agent can be considered to be fully rational because the irrational mind does not deviate from the rational mind's decision: $p(E)=p$ and $p_k(E)=p$.

A degree of decreasing rationality is determined by the range $0 \leq k < 1$. 
When $k=0.5$, $p_{k}(E) = 0.5$ irrespective of the rational mind's judgement of the probability of $E$. This reflects the situation in which the agent is caught between two minds with no means of resolving the conflict between the two, so the ultimate decision is random.
Finally, $k=0$, $p(E)=p$ and $p_k(E)=1-p$ expresses full irrationality as the irrational mind contradicts the judgement of the rational mind. Note that both $k=0$ and $k=0.5$ can be considered irrational, depending on the theory of rationality to which one might subscribe. If an agent's choice is essentially random ($k=0.5$), this may describe an internal irrationality associated with their decision. If normative measures of irrationality are considered, such that there is an objectively rational choice, then a final decision which departs completely from the rational choice ($k=0$) can define a maximum level of irrationality. Regardless, irrationality increases as values of $k$ decrease.

We do not claim that defining irrationality in the preceding way presents a complete picture.
  In fact, irrationality can be considered in different ways, e.g., in relation to ideas and theories about anti-realism \cite{hoffman2014objects} to some theories about biases and perceptions in cognitive science \cite{costello2014surprisingly,costello2016people,costello2017explaining,costello2018surprising,costello2019rationality}. 
  However, by parameterizing irrationality by a bistable parameter, this method allows degrees of irrationality to be investigated in a systematic way.\\
  \begin{figure}[]
    \centering
    \vspace{-5cm}
    \includegraphics[width=12 cm]{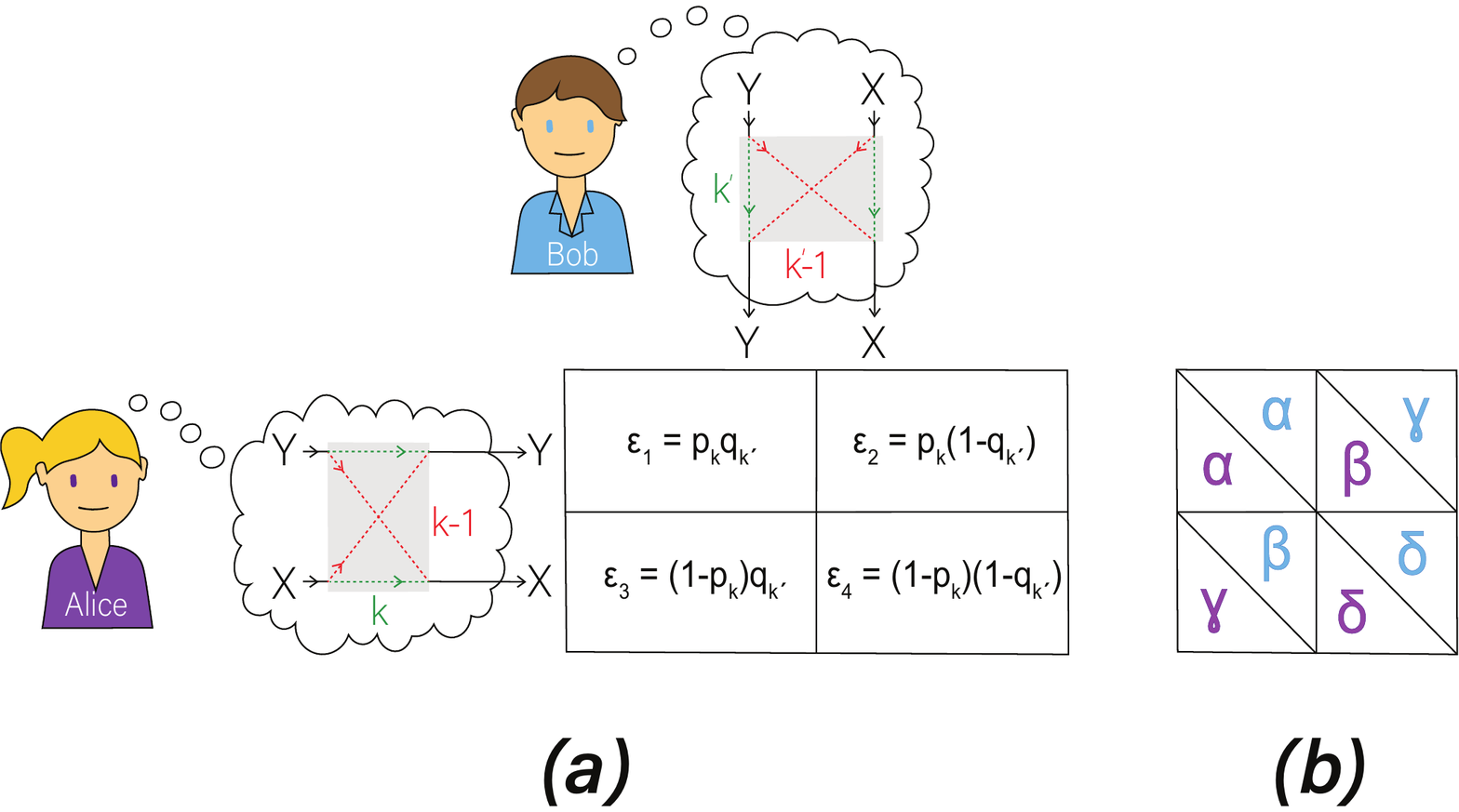}
    \vspace{-5cm}
     \caption{(a) Agents, e.g. Alice and Bob, choose rationally ideal strategies, Y and X, while bistable parameters deform these decisions to the alternative strategy with probabilities $k$ and $k^{\prime}$. Probabilities $\epsilon_{i}=\epsilon_{i}(p,q;k,k^{\prime})$, $i=1,\cdots,4$, denote probabilities of outcomes of different strategies subject to bistable parameters. (b) Alice's payoffs for choosing $(Y,Y)$, $(Y,X)$, $(X,Y)$ and $(X,X)$ are respectively $\alpha$, $\beta$, $\gamma$ and $\delta$, while for Bob are given by $\alpha$, $\gamma$, $\beta$ and $\delta$.      }
    \label{fig111}
\end{figure}
\indent Consider a game, such as the one defined in Fig. \ref{fig111} with two players, Alice and Bob. Alice and Bob must decide on a strategic option $X$ or $Y$ leading to four possible strategies after a single move $(X,X),(X,Y),(Y,X)$ and $(Y,Y)$, where the first component in the ordered pair represents Alice's decision and the second represents Bob's decision.
Ideally assuming Alice and Bob to be rational, the probability of each choosing option $Y$ is denoted by $p$ and $q$ respectively.
However, considering irrationality, a given agent may deviate from the rational choice. Bistability captures this, and the corresponding bistable probabilities are denoted $p_k$ and $q_{k^{\prime}}$ where 
the bistable parameter $k$ models the (ir)rationality of Alice's decision making and $k'$ models the (ir)rationality of Bob's decision making.
The probabilities of the four possible strategies are denoted $\epsilon_{i}=\epsilon_{i}(p,q;k,k^{\prime})$, $i=1,\cdots,4$.\\
\section{Classical bistable game}\label{sec3}
\begin{figure*}[]
\centering 
    \includegraphics[width=14cm]{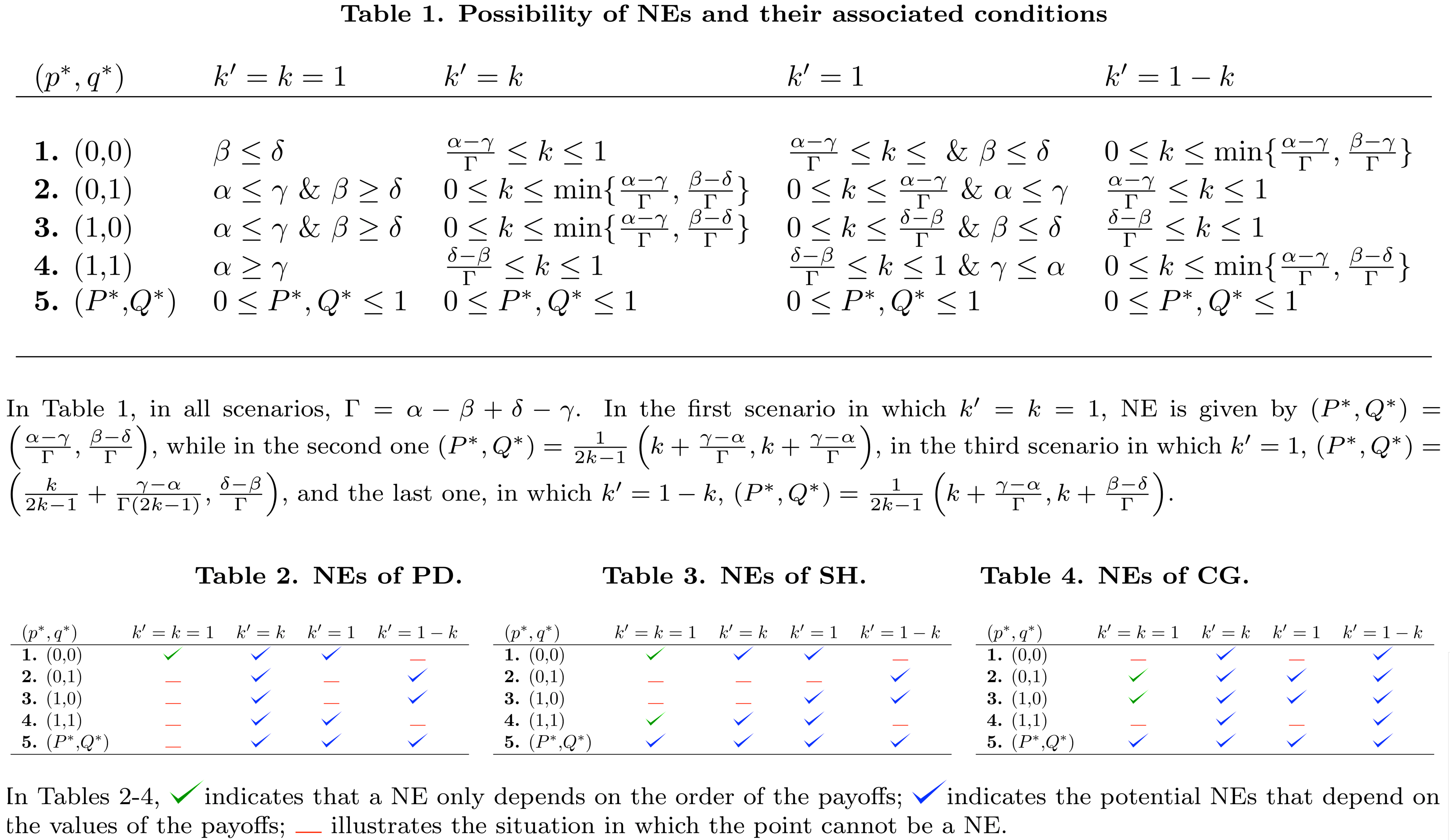}

     \caption{Table 1 illustrates all possible NEs and their associated conditions  for four different scenarios; the rational case in which $k^{\prime}=k=1$, the symmetrical irrational case, i.e., $k^{\prime}=k$ (such that $k\neq 1$), the case where one agent is irrational, $k^{\prime}=1$ and finally the case of complementary irrationality, in which $k^{\prime}=1-k$. Tables 2-4 show the NEs for respectively the PD, SH and CG.   }
    \label{tables}
\end{figure*}

We consider a scenario in which two agents, e.g. Alice and Bob, cannot communicate with each other. This is equal to no-signaling in physics \cite{brunner2013connection}.  
In any round of the game, either agent can choose one of two possible options, X and Y, as part of their strategy (i.e. X and Y in Fig. \ref{fig111}(a)).
Experimental results are recorded for a large number of individual trials so that each experimental trial comprises two choices, denoted $+1$ and $-1$. 
Payoffs are awarded depending on agent decisions over many trials, the matrix of the game they play and the statistics of the measurement outcomes.  
It turns out that when agents are influenced by the bistable parameter, the probabilities $\epsilon_{i}$'s are non-factorisable, and therefore a bistable game simulates a quantum game \cite{brunner2013connection,iqbal2016equivalence}. 
To study the impact of bistablity, three different scenarios are considered. In the first, we consider $k^{\prime}=k$, which means the level of (ir)rationality in both agents is equal. In the second, we consider situations in which one agent is irrational ($k<1$), while the other is not ($k^{\prime}=1$). 
The third scenario considers situations in which both payers are influenced by complementary levels of (ir)rationality, $k^{\prime}=1-k$. Moreover, when $k=k^{\prime}=0.5$, a NE is obtained, regardless of the payoffs. 
 The curious fact in this case is that the outcome of the game does not depend on the strategies of the respective agents. \\   
\indent In a two-agent, two-choice game, there can be up to five possible NEs based on the payoffs. Fig. \ref{tables}, Table 1. specifies the conditions which have to be fulfilled in order for NEs to be obtained. The number of NE depend on both the payoffs and probability of options chosen by the agents. Note that the first four NE, i.e. $(0,0)$, $(0,1)$, $(1,0)$ and $(1,1)$, are pure strategies, as that they can be determined in a single round. The last NE $(P^*,Q^*)$ denotes a mixed strategy, meaning that it can only be determined through a sample of plays taken from the game, and it is presented in Table \ref{tables}.  
\subsection{Prisoner's Dilemma} 
PD is an example of a non-zero sum game. Contrary to zero sum games in which agents' gains or losses are exactly balanced between them, in non-zero sum games, both agents may not necessarily be engaged and benefiting from strictly opposing strategies, but rather benefit from mutual cooperation.

In PD, each agent must independently choose between two actions: to cooperate (C) or to defect (D). In the classic version of the game, 'independent' means that there is no way for the two agents to communicate with each other, so one agent is never aware of the strategy of the other. Agents are assumed to rely on the reward (or payoff) associated with each strategy when making a decision, with the objective of maximizing their individual payoffs.

In a classical setting, the PD game has one {\it single dominant strategy}: to defect. The dilemma in the game comes from the fact that if both agents choose the dominant strategy, which is to defect, then they perform substantially worse than if they had both chosen to cooperate. In game theory, this strategy also coincides with the NE. When both agents choose to defect, then it seems that there is no other possible strategy that could bring a higher individual payoff to the agents. With the study of bistable games, one can avoid this situation where agents are attached to a single NE corresponding to the defect strategy by providing agents alternative NEs with which they can engage in their strategies.

\begin{figure*}
    \centering
    \includegraphics[width=15cm]{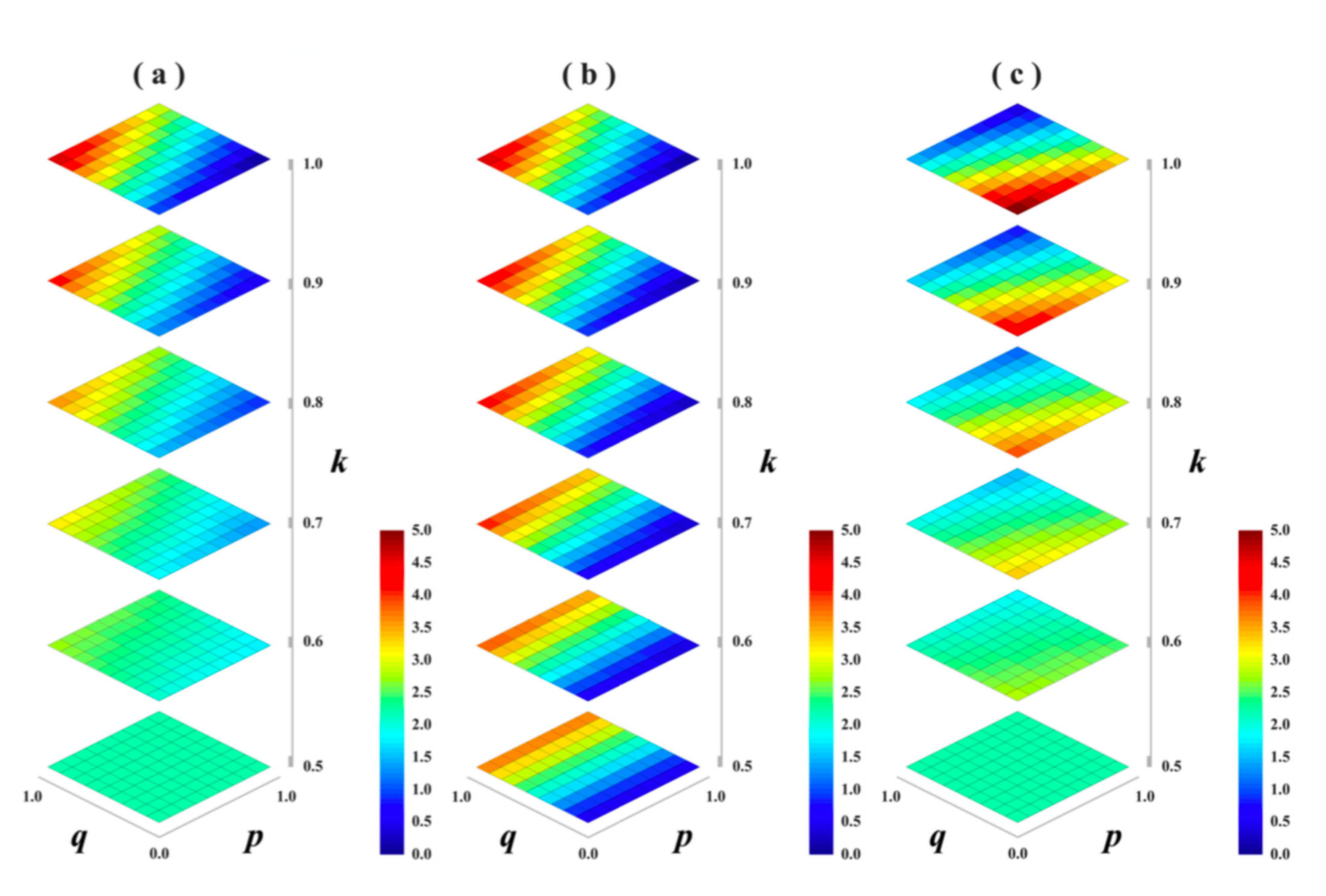}\label{figa}
     \caption{Plot (a) indicates the utility of Alice  for different values of the bistable parameter, while the bistable parameter is the same for both agents, $k=k^{\prime}$. The plot (b) presents the  utility in the case in which $k^{\prime}=1$. Finally, the plot (c) depicts the  situation in which $k=1-k^{\prime}$.
     For all plots: $\alpha=3, \beta=0, \gamma=1$ and $\delta=5$.}\label{fig1}
\end{figure*}

In a traditional PD game, a payoff matrix is associated to each of the agents' actions, as demonstrated in Fig. \ref{fig1}(b). For this specific game, the payoffs of the payers need to satisfy the following rules:  $ \delta > \alpha > \gamma > \beta $ and $2\alpha> \beta + \gamma$ \cite{press2012iterated}. To expand on the scenarios that result in these payoffs, when both agents cooperate, they both have a payoff of $\delta$, when they both defect, they have a payoff of $\alpha$ and when they engage in different strategies, the agent choosing to cooperate receives $\gamma$, while the agent who chooses to defect gets $\beta$.
In order to analyze the impact of varying degrees of irrationality, the three previously mentioned  scenarios are investigated: a) $k=k^{\prime}$, b) $k^{\prime}=1$ and c) $k^{\prime}=1-k$, where $k$ is the bistable parameter that models the (ir)rationality of Alice and $k^{\prime}$ models the (ir)rationality of Bob. 

  \indent Table (2), in Figure \ref{tables},  illustrates NEs in four different scenarios. Despite the fact that in case $k^{\prime}=k=1$, which corresponds to the usual PD where both agents are rational and where there is  only one NE, in other cases, in which at least one agent is influenced by the bistable parameter, the number of NEs increase up to the maximum of three. 
  For example, in the scenario in which $k=k^{\prime}$, based on the payoffs, either $(0,0)$, $(1,1)$ and $(P^{\ast},Q^{\ast})$  or $(0,1)$, $(1,0)$ and $(P^{\ast},Q^{\ast})$ can be NEs. In the case of $k^{\prime}=1$, based on the order of payoffs in the PD, only $(0,0)$, $(1,1)$ and $(P^{\ast},Q^{\ast})$ are potentially NEs. In the last scenario, in contrast with the PD without any bistablity, $(0,0)$ and $(1,1)$, definitely cannot be NEs. 
  
  We summarize the impacts of the bistable parameter on the PD game as the following: {\it (i)} Mixed strategy $(P^{\ast}, Q^{\ast})$ can yield a NE; {\it (ii)} The number of NEs depends on the values of the payoffs and the level of irrationality. The maximum number of NEs is three.    \\   
  %
  %
  %
%

\indent Plots in Fig. \ref{fig1} (a) show Alice's utility  in a symmetrical  bistable PD game, i.e., $k=k^{\prime}$ with $\alpha=3$, $\beta=0$, $\gamma=1$ and $\delta=5$ for different levels of irrationality for Alice $k=0.5,\cdots, 1$. 
These plots indicate that by increasing the level of the agents' irrationality, their utility becomes increasingly independent of  the rational decision with regards to a move. Finally, their utility becomes completely independent from their rational decisions when the bistable parameter equals $0.5$. 
Plots in Fig. \ref{fig1} (b) indicate Alice's utility using the same parameters but where Bob is completely rational ($k^{\prime}=1$). Drawing a comparison between plots in (a) and (b) demonstrates that Alice's utility can be at a maximum whenever Bob acts rationally, $k^{\prime}=1$.
Finally,  plots in (c)  illustrate the scenario where Alice and Bob have complementary levels of irrationality, i.e., $k=1-k^{\prime}$. Despite the fact that defection/cooperation is the NE, the maximum payoff for Alice is obtained whenever she cooperates with Bob. Also, the maximum utility is less than two previous scenarios (a) and (b).

In order to illustrate how the bistable probability framework relates to published experimental findings, we apply it to the findings in \cite{capraro2014heuristics}. 
This study utilised a continuous strategy Prisoner’s Dilemma paradigm where participants were given a choice of varying levels of cooperation, rather than a binary choice between cooperate and defect. The effect of a benefit/cost ($b/c$) ratio was investigated by varying a constant that multiplied the amount that the player’s partner gained by the amount a participant chose to donate to them. 
The study found that the majority of participants fell under three categories: maximum cooperation, maximum defection (i.e. no cooperation), and $50$\% cooperation. Increasing the $b/c$ ratio resulted in an increase to maximal cooperation and a decrease to maximal defection.\\
\begin{figure}
    \centering
\includegraphics[width=15cm]{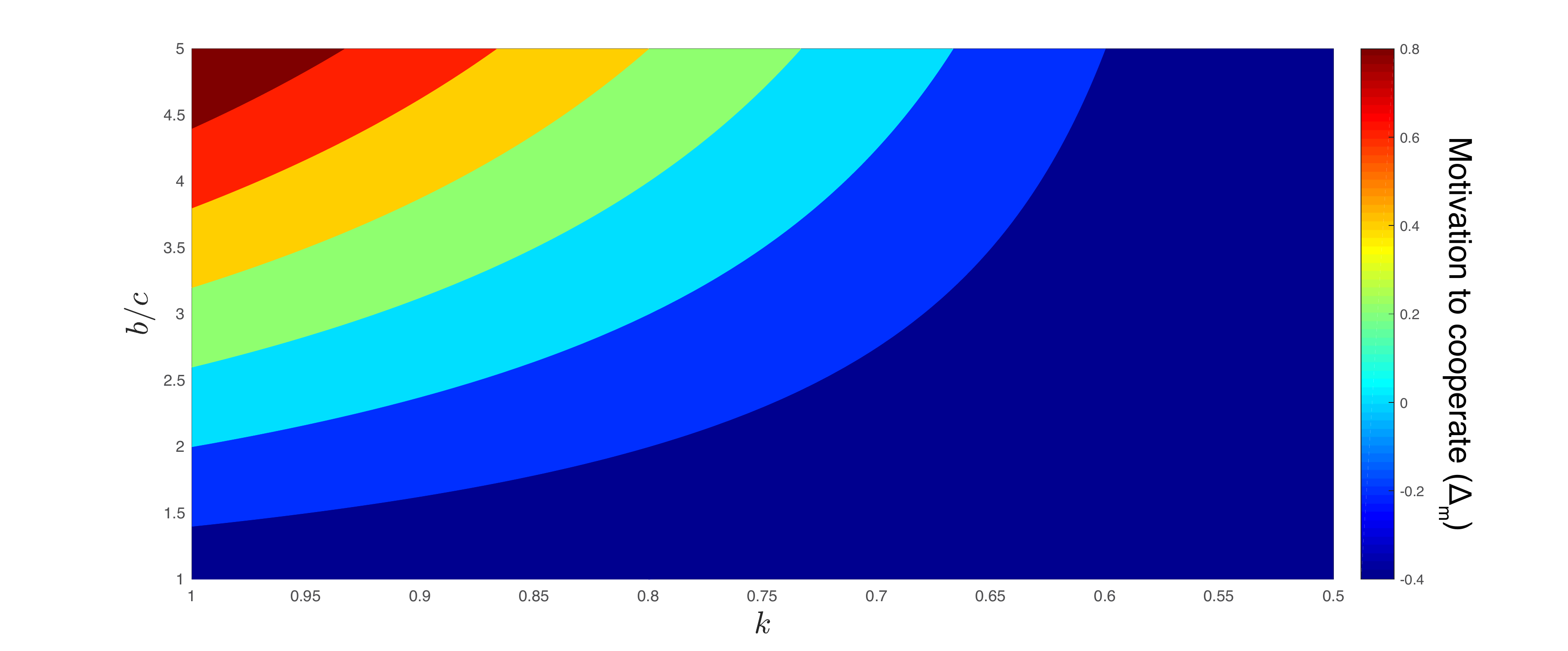}
    \caption{Motivation to cooperate $\Delta_{m}$ as a function of bistable parameter $k$ and cost-benefit ratio $b/c$.}
    \label{cooperative}
\end{figure}{}
\indent The $b/c$ ratios on motivation to cooperate were varied within the  bistable framework, and compared to the findings presented in \cite{capraro2014heuristics}.
In order to do this, the payoffs of each strategy are defined as follows:
\begin{eqnarray}
   \alpha=b-c, \ \beta= -c, \ \gamma=0, \ \delta=b,
\end{eqnarray}
which means an agent pays a cost $c$ with respects to benefit $b$ ($b>c$). 
This is therefore the difference between the gain that the agent is able to achieve:    
\begin{eqnarray}
   \Delta_{m}&=& Gain Expectation  of Cooperation-Gain Expectation of Defection\nonumber\\
   &=&\left[p_{k}q_{k^{\prime}}(b-c)-cp_{k}(1-q_{k^{\prime}})\right]
   - \left[b(1-p_{k})(1-q_{k^{\prime}})\right],
\end{eqnarray}
 which can be considered as a motivating factor for cooperation among agents.  
 In Fig. \ref{cooperative}, the motivation to cooperate $\Delta_{m}$ is illustrated as function of two parameters, i.e.,  the bistable parameter $k$ and the $b/c$ ratio.
 When $k=1$ (i.e., a player is rational) an increase in $b/c$ ratio coincides with an increased motivation to cooperate, which is consistent with results presented in \cite{capraro2014heuristics}. However, as $k$ decreases (i.e., the the deviation from rationality increases), the overall motivation to cooperate decreases, which corresponds to an increased likelihood to defect. 
 An irrational agent can therefore be seen to become less sensitive to the $b/c$ ratio, and become increasingly motivated to defect regardless of $b/c$. 
 This finding can be placed in the context of the cognitive bias of risk aversion, where agents tend to weigh risk and reward asymmetrically. Although the above does not provide a direct explanation for the subset of participants in \cite{capraro2014heuristics} who chose 50\% cooperation regardless of the $b/c$ ratio, it does illustrate an insensitivity to changes in a $b/c$ ratio for irrational players, which could be explained in terms of heuristics.
 \subsection{Stag Hunt}

SH is a prototype of a social contract between two parties, originally proposed by the philosopher Jean-Jacques Rousseau to question the degree to which people are faithful to their contracts. The traditional version of the game puts two agents in a hunting scenario. Each of the agents has the choice of pursuing either a rabbit or a stag. The stag draws the largest reward, however the rabbit is enough to satisfy an individual agent. What Jean-Jacques Rousseau questioned is that if both agents want to hunt the stag, then they need to be faithful to their posts and strategy. But if a rabbit happens to pass within an agent's reach, they face a choice to either remain at their post, or break the contract and pursue the rabbit for individual satisfaction. In this sense, the SH game suggests a scenario where the two agents can jointly cooperate to get the largest reward, or be selfish and get a satisfactory individual reward. \\
\indent The payoffs for this game conform to the following rules: $ \alpha> \gamma \geq \delta > \beta$ and $\gamma + \delta >\alpha+\beta$, where $\alpha$ is the maximum payoff that Alice and Bob can obtain in a mutual cooperation strategy, $\gamma$  corresponds to the payoff when Alice decides to break the contract and pursue the rabbit, $\delta$ is the payoff when Bob decides to break the contract and pursue the rabbit, and finally, $\beta$ is the payoff that Alice and Bob both obtain if they both break the contract and decide to hunt the rabbit. \\
\indent The order of payoffs in this game leads to two NE states: (1) both agents choose to hunt the stag and (2) both agents choose to hunt the rabbit. By exploring bistable parameters, one can find additional strategies that can release agents from these two NE for means of engaging in alternative strategies.\\
\indent  Table (3), in Fig. \ref{tables} represents NEs in the three different scenarios.  In the first scenario a) in which $k=k^{\prime}=1$, there are up to three different NEs. In fact, two pure strategies $(0,0)$ and $(1,1)$ are always fulfilled by NE inequalities,  while the mixed strategy $(P^{\ast},Q^{\ast})$ as the third NE depends on values of payoffs. In the second  scenario, in which $k=k^{\prime}\neq 1$, there is a maximum of three NEs.
   In the scenario b) in which  $k^{\prime}=1$, except the point $(0,1)$, all other points can potentially yield NE, based on specific values of payoffs. In the last scenario c), there is a maximum of three NEs, namely points $(0,1)$, $(1,0)$ and $(P^{\ast},Q^{\ast})$. 

 \subsection{Chicken Game}
 

CG is another type of game that models coordination strategies between two agents. The game describes two agents who are heading towards each other. If one of the agents deviates their path in order to avoid a collision, then this agent loses and is labeled the 'chicken' (implied as the defeated), while the other is considered the victor. 

Under this scenario, the strategies for NEs are straightforward. If Alice keeps heading straight and Bob swerves, then Alice is engaged in a dominant strategy where, regardless of the action she chooses, it will not bring a higher payoff. Consequently, this obtains a NE. The same NE can be found if Bob continues straight and Alice deviates. 
The payoffs for this game have the following order: $\gamma>\alpha>\beta\gg \delta$ subject to  the following condition: $0 < \beta < (\gamma+\beta)$. Under this order, the parameters $\gamma$ and $\alpha$ respectively correspond to the payoffs of Alice and Bob when they engage in the dominant strategy.



\indent In the first scenario a), in which $k=k^{\prime}=1$, there are maximum up to three NEs. 
In this scenario, while differences in payoffs matter for calculating the mixed strategy NE, i.e. $(P^{\ast},Q^{\ast})$, only the order of payoff matters for determining regular NEs, i.e., $(0,1)$ and $(1,0)$. In the second scenario b), $k=k^{\prime}\neq 1$, there is a maximum of three NEs. Based on the payoffs, either  points $(0,0)$, $(1,1)$ and $(P^{\ast},Q^{\ast})$ or $(0,1)$, $(1,0)$ and $(P^{\ast},Q^{\ast})$ are NEs. In the scenario which $k^{\prime}=1$, there is a maximum of three NEs, depending on the payoffs.  In the last scenario c) in which $k^{\prime}=1-k$, subject to payoffs, up to three NEs are determined, i.e.,  points  $(0,1)$, $(1,0)$, and $(P^{\ast},Q^{\ast})$, or $(0,0) $, $(1,1)$ and $(P^{\ast},Q^{\ast})$.

\section{Quantum Bistable games}\label{sec4}
 Quantum game theory has been suggested as a way to generalize game theory by exploiting concepts inherent to quantum theory such as entanglement. \cite{meyer1999quantum,eisert1999quantum,eisert2000quantum}.
Quantum games are based on each agent having a qubit. 
A quantum move in the game is modelled by applying a unitary matrix to the given agent's qubit.
If agents’ qubits become entangled, there are hidden channels of communication. 
Due to the fact that accessible information in a quantum system is greater than in a classical system \cite{pitowsky1994george,goh2018geometry,vourdas2019probabilistic},  a quantum game can feature more strategies. One potential result of this fact is  that the number of NEs can differ from the classical variant of the game. 
Notwithstanding the extra features that quantum games provide, the quantum game-theoretic framework still assumes rationality, e.g. agents focus on maximizing their expected utility.
\noindent
We begin by observing that the utilities of agents in a classical game can be formalized as expectation values of sharp projections when the initial states of agents are modelled by a two dimensional real Hilbert space. 
This observation opens the door to generalise strategies in classical games by means of unitary operators. 
A second observation is that unsharp projections as positive-operator values (POV) projections is a means to generalize measurement in quantum mechanics.  
\noindent
 Both observations allow irrational quantum games to be modelled. \\
\indent In these games, Bob's and Alice's strategies are defined by  unitary operators in a complex Hilbert space and their associated expected utilities are given by  expectation values computed by means of bistable Kraus (POV) projections. 
We assume that Alice's and Bob's initial states are described by a 
a Bell state of the form $|\psi_{i}\rangle=(|0,0\rangle+i|1,1\rangle)\sqrt{2}$, which is a maximally entangled state.
In this section, quantum bistable variants of PD, SH and CG are studied. 
 \begin{figure*}
    \centering
    \includegraphics[width=15 cm]{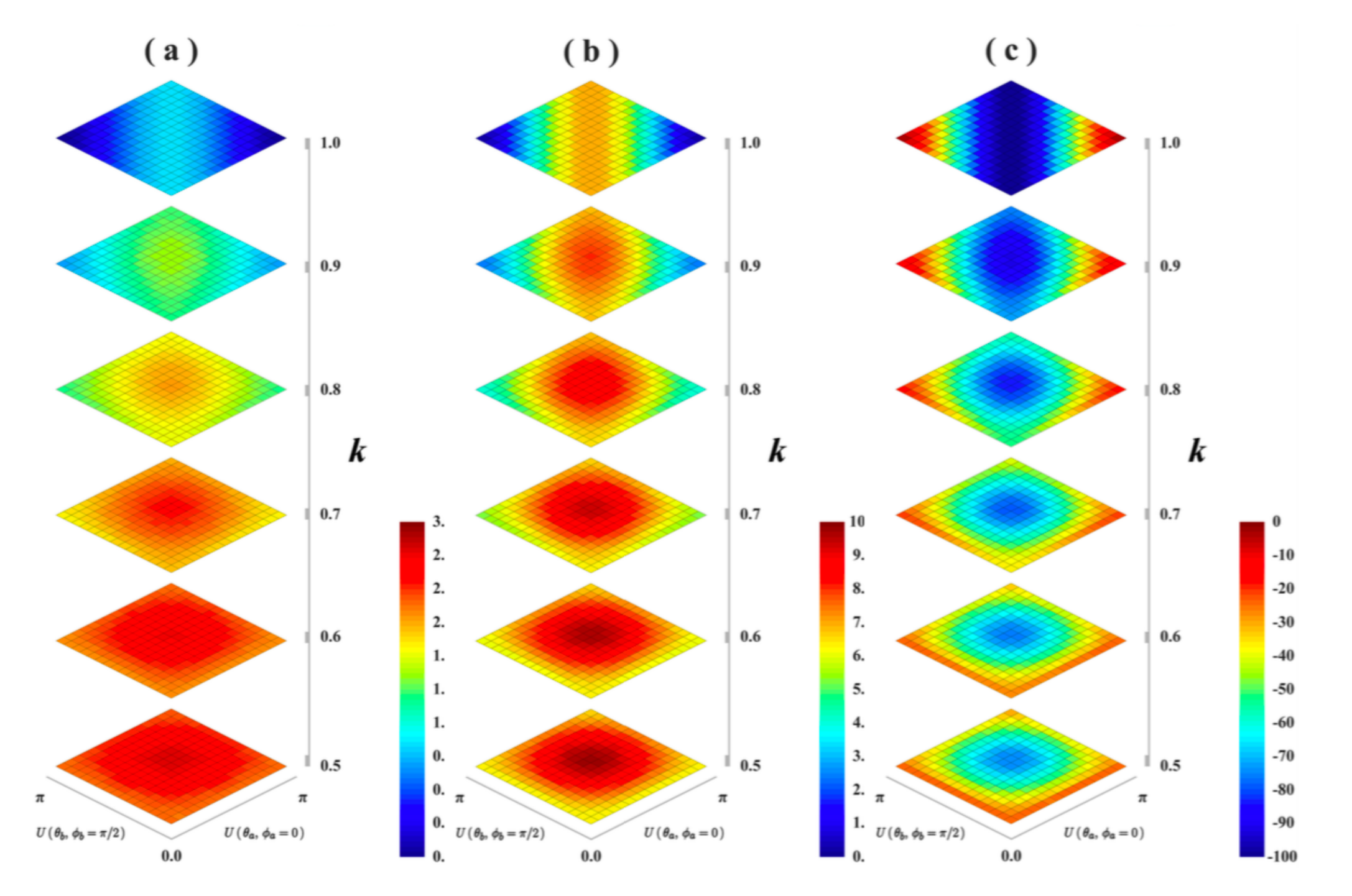}
      \caption{Utility functions of Alice  for different operators $\hat{U}(\theta_{i},\phi_{i}=0)$, $i= a, b$, are plotted in plots (a), (b) and (c) for different games, respectively PD, ST and CG. }
    \label{pdqu}
\end{figure*}
\subsection{Prisoner's Dilemma}
In this subsection, we study the effect of  bistable parameters $k$ and $k^{\prime}$ on the quantum PD game. 
When $0.5 < k=k^{\prime}\leq 1$, the point $(\theta_{a}^{\ast},\theta_{b}^{\ast})=(\pi/2,\pi/2)$ corresponds to a NE.
When
$0.5 < k=1-k^{\prime} \leq 1$, the point $(\theta_{a}^{\ast},\theta_{b}^{\ast})=(0,0)$ is a NE.\\
\indent Also, we study the effects of irrationality on the utility  of agents. Plots (a) in Fig. \ref{pdqu} demonstrate the utility function of Alice for different values of bistability when Alice and Bob are equally irrational, i.e., $k=k^{\prime}$. 
These plots illustrate that increasing the bistable parameter causes the utility to increase.
In other words, the more irrational Alice becomes, the greater her utility.
Recall in the classical variant of the bistable game when $k=k^{\prime}=0.5$ the utilities of Alice and Bob are independent of their strategies.
However, in the quantum variant the strategies of the agents do have an effect on their utilities. In Alice's case, this effect is due to the initial entangled state and the phase factor $\phi_{a}$ associated with her strategy.
\subsection{Stag Hunt}
In the case of $0.5<k=k^{\prime}\leq 1$, $(\theta_{a}^{\ast},\theta_{b}^{\ast})=(\pi/2,\pi/2)$ and $(\theta_{a}^{\ast},\theta_{b}^{\ast})=(0,0)$ are two pure strategies that yield NEs,  while $(\theta_{a}^{\ast},\theta_{b}^{\ast})=(\cos^{-1}\sqrt{(\delta-\beta)/\Gamma},\cos^{-1}\sqrt{(\delta-\beta)/\Gamma})$ is a mixed NE. In addition, The scenario in which $0.5<k=1-k^{\prime}\leq 1$, points $(\pi/2,0)$, $(0,\pi/2)$ and $(\cos^{-1}\sqrt{(\delta-\beta)/\Gamma},\cos^{-1}\sqrt{(\delta-\beta)/\Gamma})$ are NEs with $\Gamma = \alpha - \beta + \delta - \gamma$. Plots in Fig. \ref{pdqu} (b) demonstrate the effect of the bistable parameter $k=k^{\prime}$ on Alice's utility. Drawing a comparison between plots in (b), indicates maximum utility is obtained when both agents' strategies approach $(\pi/2,\pi/2)$. 
Like the previous case when Alice and Bob were equally irrational, the level of utility increases in accordance with increased levels of irrationality i.e., $k$ decreases.

Finally, when $k=k^{\prime}=0.5$ the utility depends on the respective  strategies, which is in direct contrast to the same scenario in the classical bistable game, where utility was shown to be independent of the given strategy. 
\subsection{Chicken Game} 
When $0.5<k=1-k^{\prime}\leq 1$ the quantum bistable CG
yields up to three NEs, i.e., $(\theta_{a}^{\ast},\theta_{b}^{\ast})=(0,\pi/2)$, $(\theta_{a}^{\ast},\theta_{b}^{\ast})=(\pi/2,0)$ and $(\theta_{a}^{\ast},\theta_{b}^{\ast})=(\cos^{-1}\sqrt{(\delta-\beta)/\Gamma},\cos^{-1}\sqrt{(\delta-\beta)/\Gamma})$.

When another scenario is considered in which $0.5<k=1-k^{\prime}\leq 1$, up to three NEs can be found, but the  pure NEs turn out to differ from the previous case, namely $(0,0)$, and $(\pi/2,\pi/2)$, while the mixed NE is similar with the previous scenario. Fig. \ref{pdqu}-Plots in (c) illustrates the impact of the bistable parameter on the utility. 
Once again,  the level of utility increases in accordance with increased levels of irrationality i.e., $k$ decreases.
 Also, the maximum utility for Alice is obtained whenever one of the agents' strategies is determined by an operator in the $z$-direction, i.e., $\theta=0$.

\section{Methods}\label{sec5}
The following provides the theoretical details of both classical and quantum bistable games.
%
\subsection{Nash equilibria for a general classical bistable game}
Utilities for Alice and Bob are specified as follows:
\begin{eqnarray}
\Pi_{A}(p,q;k)&=& \alpha \epsilon_{1} +\beta \epsilon_{2} + \gamma \epsilon_{3} + \delta \epsilon_{4}, \label{g2.2}\\
\Pi_{B}(p,q;k)&=& \alpha \epsilon_{1} +\gamma \epsilon_{2} + \beta \epsilon_{3} + \delta \epsilon_{4}. \label{g2.3}
\end{eqnarray}
in which
 $   \epsilon_{1}=p_{k} q_{k^{\prime}}$,  $\epsilon_{2}=p_{k}(1-q_{k^{\prime}})$, 
$ \epsilon_{3}=(1-p_{k})q_{k^{\prime}}$ and
    $\epsilon_{4}=(1-p_{k})(1-q_{k^{\prime}})$.
It is interesting that the  probability distributions  $\epsilon_{i}$'s are naturally non-factorizable and can therefore generate a quantum-like game \cite{brunner2013connection,iqbal2016equivalence}.
According to the definition of NE, i.e.,
\begin{eqnarray}
    \Pi_{A}(p^{\ast},q^{\ast})-\Pi_{A}(p^{\ast},q)\geq 0,\hspace{.2cm}
    \Pi_{B}(p^{\ast},q^{\ast})-\Pi_{B}(p,q^{\ast})\geq 0,
\end{eqnarray}
and using relations (\ref{g2.2}) and (\ref{g2.3}),  the Nash strategies  are given by 
\begin{eqnarray}
(2k^{\prime}-1)(q^{\ast}-q)\Big[\left(\alpha-\beta+\delta-\gamma\right)p_{k}^{\ast}+\beta-\delta
\Big]\geq 0,\label{eq2.5}\\
(2k-1)(p^{\ast}-p)
\Big[\left(\alpha-\beta+\delta-\gamma\right)q_{k^{\prime}}^{ \ast}+\beta-\delta
\Big]\geq 0.\label{eq2.6}
\end{eqnarray}
\subsection{Nash Equilibria for a general quantum bistable game}  
In bistable quantum games Bob's and Alice's strategies are modelled by a unitary operator. 
Their irrational decisions ($k <1$) are defined by a bistable projection operator:
\begin{eqnarray}\label{eq4.1}
    \mathbb{P}_{\pm}(\hat{n};k)=(1-k)\mathbb{I}_{2\times 2}+      (2k-1)\pi_{\pm}^{\hat{n}}
\end{eqnarray}
in which $k$ is the bistable parameter in the interval $0\leq k\leq 1$, $\mathbb{I}_{2\times 2}$ and $\pi_{\pm}^{\hat{n}}$ are respectively the identity matrix and projection operator in the $\hat{n}$-direction,
\begin{eqnarray}\nonumber
    \mathbb{I}_{2\times 2}&=&\left(
    \begin{array}{cc}
        1 & 0 \\
        0 & 1
    \end{array}
    \right), \\
    \pi_{+}^{\hat{n}}&=&\left(
    \begin{array}{cc}
        \cos^{2}\frac{\vartheta}{2} & e^{-i\varphi}\sin \frac{\vartheta}{2} \cos \frac{\vartheta}{2} \\
        e^{i\varphi}\sin \frac{\vartheta}{2} \cos \frac{\vartheta}{2} & \sin^{2}\frac{\vartheta}{2}
    \end{array}
    \right),\\
    \pi_{-}^{\hat{n}}&=&\left(
    \begin{array}{cc}
        \sin^{2}\frac{\vartheta}{2} & -e^{-i\varphi}\sin \frac{\vartheta}{2} \cos \frac{\vartheta}{2} \\
        -e^{i\varphi}\sin \frac{\vartheta}{2} \cos \frac{\vartheta}{2} & \cos^{2}\frac{\vartheta}{2}
    \end{array}
    \right),
\end{eqnarray}
where $\hat{n}=(\sin \vartheta \cos \varphi, \sin \vartheta \sin \varphi, \cos \vartheta )$. Note that in quantum mechanics, the bistable  parameter $k$ (which in physics might be analogous with a parameter to model noise\footnote{Note that this analogy is imprecise as we do not expect the effect of noise to be comparable with the main system.}, see for example \cite{liang2011specker}) is defined by a number between $0.5$ and $1$. \\
\indent Now, if we consider the bistable projections in direction of $z$-axis,
\begin{eqnarray}\label{eq9}
    \mathbb{P}_{+}(\hat{z};k)=\left(
    \begin{array}{cc}
    k     & 0 \\
        0 & 1-k
    \end{array}\right),\ 
    \mathbb{P}_{-}(\hat{z};k)=\left(
    \begin{array}{cc}
    1-k     & 0 \\
        0 & k
    \end{array}\right).
\end{eqnarray}
by considering a binary distribution, i.e., $|\psi\rangle=(\sqrt{p},\sqrt{1-p})^{T}$, the probability of finding any option is obtained by expectation values of the relations (\ref{eq9}),  $\langle\psi | \mathbb{P}_{\pm}(\hat{z};k)|\psi\rangle=P_{k}(\pm)$, i.e. 
\begin{eqnarray}
    P_{k}(+)&=&1-p-k+2kp\nonumber\\
    &=&\left(\sqrt{p} \ \sqrt{1-p}\right) \left(
    \begin{array}{cc}
    k     & 0 \\
    0     & 1-k
    \end{array}
    \right)\left(
    \begin{array}{c}
         \sqrt{p}  \\
          \sqrt{1-p}
    \end{array}
    \right),\\
    P_{k}(-)&=&p+k-2kp\nonumber\\
    &=&\left(\sqrt{p} \ \sqrt{1-p}\right) \left(
    \begin{array}{cc}
    1-k     & 0 \\
    0     & k
    \end{array}
    \right)\left(
    \begin{array}{c}
         \sqrt{p}  \\
          \sqrt{1-p}
    \end{array}
    \right).
\end{eqnarray}

\indent We consider a quantum game in which the initial state is maximally entangled, i.e., a Bell state, $|\psi_{i}\rangle=(|0,0\rangle+i|1,1\rangle)\sqrt{2}$. 
Alice and Bob choose their own quantum strategy which is defined by the following unitary operator,
\begin{eqnarray}
    \hat{U}=\hat{U}_{A} \otimes \hat{U}_{B}
\end{eqnarray}
in which
\begin{eqnarray}
    U_{A}&=& \left(
    \begin{array}{cc}
    e^{i\phi_{a}}\cos\frac{\theta_{a}}{2}     & \sin\frac{\theta_{a}}{2} \\
       - \sin\frac{\theta_{a}}{2} & e^{-i\phi_{a}}\cos\frac{\theta_{a}}{2}
    \end{array}
    \right), \nonumber\\
    U_{B}&=& \left(
    \begin{array}{cc}
    e^{i\phi_{b}}\cos\frac{\theta_{b}}{2}     & \sin\frac{\theta_{b}}{2} \\
       - \sin\frac{\theta_{b}}{2} & e^{-i\phi_{b}}\cos\frac{\theta_{b}}{2}
    \end{array}
    \right),\nonumber
\end{eqnarray}
where $\theta_{a},\theta_{b}\in [0,\pi]$ and $\phi_{a},\phi_{b}\in [0,\pi/2]$. Hence, the final state is given by $|\psi_{f}\rangle=\hat{U} |\psi_{i}\rangle $, i.e., $|\psi_{f}\rangle=\left(\begin{array}{cccc}
    \psi_{1} & \psi_{2} & \psi_{3} & \psi_{4}
\end{array}\right)^{T}/\sqrt{2}$, in which
\begin{eqnarray}
    \psi_{1}&=&e^{i(\phi_{a}+\phi_{b})} \cos \frac{\theta_{a}}{2} \cos \frac{\theta_{b}}{2} +i    \sin \frac{\theta_{a}}{2} \sin \frac{\theta_{b}}{2}, \nonumber\\
    \psi_{2}&=&-e^{i\phi_{a}} \cos \frac{\theta_{a}}{2} \sin \frac{\theta_{b}}{2} +i e^{-i\phi_{b}}   \sin \frac{\theta_{a}}{2} \cos \frac{\theta_{b}}{2}, \nonumber\\ 
    \psi_{3}&=&-e^{i\phi_{b}} \sin \frac{\theta_{a}}{2} \cos \frac{\theta_{b}}{2} +ie^{-i\phi_{a}}\cos \frac{\theta_{a}}{2} \sin \frac{\theta_{b}}{2}, \nonumber\\
    \psi_{4}&=& \sin \frac{\theta_{a}}{2} \sin \frac{\theta_{b}}{2} +i e^{-i(\phi_{a}+\phi_{b})}   \cos \frac{\theta_{a}}{2} \cos \frac{\theta_{b}}{2}.\nonumber
\end{eqnarray}
By using bistable projection (\ref{eq4.1}), bistable Kraus projections are given by:
\begin{eqnarray}
     \mathbb{P}_{cc}&=&\frac{1}{2} \left(
   \begin{array}{cccc}
    a_{1}    & 0 & 0 &-ia_{3} \\
    0    &  a_{2} & 0 & 0 \\
    0 & 0 & a_{2} & 0 \\
    ia_{3} & 0 & 0 & a_{1} 
   \end{array}
   \right)
   \mathbb{P}_{cd}=\frac{1}{2} \left(
   \begin{array}{cccc}
    a_{2}    & 0 & 0 &0 \\
    0    &  a_{1} & ia_{3} & 0 \\
    0 & -ia_{3} & a_{1} & 0 \\
    0 & 0 & 0 & a_{2} 
   \end{array}
   \right)\\
   \mathbb{P}_{dc}&=&\frac{1}{2} \left(
   \begin{array}{cccc}
    a_{2}    & 0 & 0 &0 \\
    0    &  a_{1} & -ia_{3} & 0 \\
    0 & ia_{3} & a_{1} & 0 \\
    0 & 0 & 0 & a_{2} 
   \end{array}
   \right)
    \mathbb{P}_{dd}=\frac{1}{2} \left(
   \begin{array}{cccc}
    a_{1}    & 0 & 0 &ia_{3} \\
    0    &  a_{2} & 0 & 0 \\
    0 & 0 & a_{2} & 0 \\
    -ia_{3} & 0 & 0 & a_{1} 
   \end{array}
   \right)
\end{eqnarray}
in which 
\begin{eqnarray}
   a_{1}&=&kk^{\prime}+(1-k)(1-k^{\prime})\\
   a_{2}&=&k^{\prime}(1-k)+k(1-k^{\prime})\\
   a_{3}&=& (2k-1)(2k^{\prime}-1) 
\end{eqnarray}
Note that by considering $k=k^{\prime}=1$ in the preceding relations i.e., Alice and Bob are rational, Kraus projections are respectively given by:
\begin{eqnarray}
    \pi_{cc}=|\psi_{cc}\rangle\langle \psi_{cc}|, \hspace{.5cm} |\psi_{cc}\rangle=\frac{|0\rangle_{a}\otimes|0\rangle_{b}+i|1\rangle_{a}\otimes|1\rangle_{b}}{\sqrt{2}},\label{eq4.4}\\
     \pi_{cd}=|\psi_{cd}\rangle\langle \psi_{cd}|, \hspace{.5cm} |\psi_{cd}\rangle=\frac{|0\rangle_{a}\otimes|1\rangle_{b}-i|1\rangle_{a}\otimes|0\rangle_{b}}{\sqrt{2}},\label{eq4.5}\\
      \pi_{dc}=|\psi_{dc}\rangle\langle \psi_{dc}|, \hspace{.5cm} |\psi_{dc}\rangle=\frac{|1\rangle_{a}\otimes|0\rangle_{b}-i|0\rangle_{a}\otimes|1\rangle_{b}}{\sqrt{2}},\label{eq4.6}\\
      \pi_{dd}=|\psi_{dd}\rangle\langle \psi_{dd}|, \hspace{.5cm} |\psi_{dd}\rangle=\frac{|1\rangle_{a}\otimes|1\rangle_{b}+i|0\rangle_{a}\otimes|0\rangle_{b}}{\sqrt{2}}.\label{eq4.7}
\end{eqnarray}
The final state $|\psi\rangle_{f}$ together with bistable Kraus projections define the agent's utilities as  follows: 
\begin{eqnarray}
    \Pi_{A}(\hat{U}_{a},\hat{U}_{b};k)&=&\alpha \langle \psi_{f}|\mathbb{P}_{cc} |\psi_{f}\rangle+\beta \langle \psi_{f}|\mathbb{P}_{cd} |\psi_{f}\rangle
    +\gamma \langle \psi_{f}|\mathbb{P}_{dc} |\psi_{f}\rangle+\delta\langle \psi_{f}|\mathbb{P}_{dd} |\psi_{f}\rangle\\
      \Pi_{B}(\hat{U}_{a},\hat{U}_{b};k)&=&\alpha \langle \psi_{f}|\mathbb{P}_{cc} |\psi_{f}\rangle+\gamma \langle \psi_{f}|\mathbb{P}_{cd} |\psi_{f}\rangle
      +\beta \langle \psi_{f}|\mathbb{P}_{dc} |\psi_{f}\rangle
      +\delta\langle \psi_{f}|\mathbb{P}_{dd} |\psi_{f}\rangle
\end{eqnarray}
in which
\begin{eqnarray}
    \langle \psi_{f}|\mathbb{P}_{cc} |\psi_{f}\rangle&=&\cos^{2}\frac{\theta_{a}}{2}\cos^{2}\frac{\theta_{b}}{2}\cos^{2}
    (\phi_{a}+\phi_{b})\nonumber\\
    &+&\frac{k+k^{\prime}-2kk^{\prime}}{2} \Big[\sin^{2}\frac{\theta_{a}}{2}\nonumber\\
    &+&\cos^{2}\frac{\theta_{a}}{2}\left(1-4\cos^{2}\frac{\theta_{b}}{2}\cos^{2}(\phi_{a}+\phi_{b})\right)\Big],\\
    \langle \psi_{f}|\mathbb{P}_{cd} |\psi_{f}\rangle&=&\cos^{2}\frac{\theta_{a}}{2}\sin^{2}\frac{\theta_{b}}{2}\cos^{2}\phi_{a} + \sin^{2}\frac{\theta_{a}}{2}\cos^{2}\frac{\theta_{b}}{2}\sin^{2}\phi_{b}\nonumber\\
    &-&2\sin\theta_{a}\sin\theta_{b}\sin\phi_{a}\cos\phi_{b}\nonumber\\
     &+&\frac{k+k^{\prime}-2kk^{\prime}}{2}
     \Big[\sin^{2}\frac{\theta_{a}}{2} \left(1-4\cos^{2}\frac{\theta_{b}}{2}
     \sin^{2}\phi_{b}\right)\nonumber\\
     &+&\cos^{2}\frac{\theta_{a}}{2} \left(1-4\sin^{2}\frac{\theta_{b}}{2}\cos^{2}\phi_{a}\right)\nonumber\\
     &+&2\sin\theta_{a}\sin\theta_{b}\sin\phi_{b}\cos\phi_{a}\Big],\\
     \langle \psi_{f}|\mathbb{P}_{dc} |\psi_{f}\rangle&=&\cos^{2}\frac{\theta_{a}}{2}\sin^{2}\frac{\theta_{b}}{2}\sin^{2}\phi_{a} + \sin^{2}\frac{\theta_{a}}{2}\cos^{2}\frac{\theta_{b}}{2}\cos^{2}\phi_{b}\nonumber\\
     &-&2\sin\theta_{a}\sin\theta_{b}\sin\phi_{a}\cos\phi_{b}\nonumber\\
     &+&\frac{k+k^{\prime}-2kk^{\prime}}{2}
     \Big[\cos^{2}\frac{\theta_{a}}{2} \left(1-4\sin^{2}\frac{\theta_{b}}{2}
     \sin^{2}\phi_{a}\right)\nonumber\\
     &+&\sin^{2}\frac{\theta_{a}}{2} \left(1-4\cos^{2}\frac{\theta_{b}}{2}\cos^{2}\phi_{b}\right)\nonumber\\
     &+&2\sin\theta_{a}\sin\theta_{b}\sin\phi_{b}\cos\phi_{a}\Big],\\
      \langle \psi_{f}|\mathbb{P}_{dd} |\psi_{f}\rangle&=&\cos^{2}\frac{\theta_{a}}{2}\cos^{2}\frac{\theta_{b}}{2}\sin^{2}
    (\phi_{a}+\phi_{b})+\sin^{2}\frac{\theta_{a}}{2}\sin^{2}\frac{\theta_{b}}{2}\nonumber\\
    &+&\frac{1}{2}\sin \theta_{a}\sin\theta_{b}\sin(\phi_{a}+\phi_{b})\nonumber\\
    &+&\frac{k+k^{\prime}-2kk^{\prime}}{2} \Big[\sin^{2}\frac{\theta_{a}}{2}\left(1-4\sin^{2}\frac{\theta_{b}}{2}\right)
    \nonumber\\
    &+&\cos^{2}\frac{\theta_{a}}{2}\left(1-4\cos^{2}\frac{\theta_{b}}{2}\sin^{2}(\phi_{a}+\phi_{b})\right)\nonumber\\
     &-& \sin \theta_{a} \sin \theta_{b}\sin (\phi_{a}+\phi_{b})\Big]
\end{eqnarray}
According to definition, the NEs for the quantum bistable game are obtained from the following inequalities:
\begin{eqnarray}
    \Pi_{A}(\hat{U}_{A}^{\ast},\hat{U}_{B}^{\ast};k)-\Pi_{A}(\hat{U}_{A},\hat{U}_{B}^{\ast};k)\geq 0, \\
    \Pi_{B}(\hat{U}_{A}^{\ast},\hat{U}_{B}^{\ast};k)-\Pi_{B}(\hat{U}_{A}^{\ast},\hat{U}_{B};k)\geq 0.
\end{eqnarray}
Explicitly, in the case in which $\phi_{a}=\phi_{b}=0$, the potential NEs are given by
\begin{eqnarray}
   f(k,k^{\prime}) \left(\cos^{2} \frac{\theta_{a}^{\ast}}{2}-\cos^{2}
    \frac{\theta_{a}}{2}\right)
    \left[\left(\alpha-\beta+\delta-\gamma
    \right)
    \cos^{2}\frac{\theta_{b}^{\ast}}{2}+\beta-\delta\right]\geq 0, \label{eqg319}\\
   f(k,k^{\prime}) \left(\cos^{2} \frac{\theta_{b}^{\ast}}{2}-\cos^{2}
    \frac{\theta_{b}}{2}\right)
    \left[\left(\alpha-\beta+\delta-\gamma
    \right)
    \cos^{2}\frac{\theta_{a}^{\ast}}{2}+\beta-\delta\right]\geq 0,\label{eqg320}
\end{eqnarray}
in which $f(k,k^{\prime})=1-2\left(k+k^{\prime}-2k k^{\prime}\right)$. As is evident,  the NE relations are independent of the bistable parameters $k$ and $k^{\prime}$ .
Moreover,  in the case in which $k=k^{\prime}=1/2$, the NE relations are in fact independent of any choices made by Bob or Alice.

\section{Conclusion }\label{sec6}
In this article, the irrationality of agents in both classical and quantum game theory has been modelled using bistable probabilities which allow irrationality to be modelled as deviations from a rational choice.
A bistable probability framework for both classical and quantum games was presented using positive operator value (POV) projections borrowed from quantum mechanics, which are used to model noise. 
However, in the framework presented in this article, the bistable parameter should not be construed as noise, but as a means of systematically investigating deviations from rationality.
This general framework was used to theoretically analyse classical and quantum variants of three well known games:  Prisoner's Dilemma, Stag Hunt and Chicken.
In each of these, Nash equilibria were shown to be a function of bistability, i.e., the level of irrationality of an agent.
As would be expected, the utilities of the payers in classical games decreased as the level of irrationality increased. \\
\indent To specify the outcomes of quantum strategies in a range of quantum games, we introduced bistable Kraus projections as a general formalism. A surprising result was seen in quantum games where the initial states of agents are maximally entangled and strategies include a phase factor. In contrast to the results seen in classical bistable games, the utility increases as the level of irrationality increases. Whenever the phase factor is excluded, calculations revealed that  NEs in irrational quantum games mimic those in rational classical games and are independent  of  the bistability  parameter, i.e., the degree of deviation from a rational choice. In addition, it is worthwhile to mention that application of quantum games in cognitive science have been generalized to contextuality scenarios, for example see  \cite{makowski2011transitivity,makowski2014cut,makowski2015transitive,makowski2017profit}, in which intransitive and transitive preferences have been studied. 
In fact, the bistable framework exhibits some potential to model contextuality \cite{dehdashti2020irrationality} and future work will be directed to the modelling of contextuality in games.
Another future direction involves extending our quantum treatment of incentivised games beyond two agents which opens the possibility of modelling agents with varying levels of entanglement.  

 
\indent 
 In summary, by defining irrationality as a deviation from a rational choice, we provide a more realistic model of decision-making in a range of game situations because both rationality and irrationality can be systematically examined in a general, unified setting. 









%
%
%


\bibliographystyle{unsrt} 
\bibliography{references1}

\end{document}